\documentclass[aps,pre,twocolumn,showpacs]{revtex4-1}

\usepackage[T1]{fontenc}
\usepackage{graphicx}
\usepackage{amsmath}

\begin{document}

\title{Model of hard spheroplatelets near a hard wall}

\author{A. Kapanowski}

\author{M. Abram}


\affiliation{Institute of Physics, Jagiellonian University, 
ulica Reymonta 4, 30-059 Krak\'{o}w, Poland}

\date{\today}

\begin{abstract}
A system of hard spheroplatelets near an impenetrable wall 
is studied in the low-density Onsager approximation.
Spheroplatelets have optimal shape between rods and plates,
and the direct transition from the isotropic to biaxial nematic
phase is present.
A simple local approximation for the one-particle distribution
function is used.
Analytical results for the surface tension and the entropy 
contributions are derived.
The density and the order-parameter profiles near the wall are calculated.
The preferred orientation of the short molecule axes is perpendicular
to the wall. Biaxiality close to the wall can appear only if the phase
is biaxial in the bulk.
\end{abstract}


\maketitle

\section{Introduction}
\label{sec:intro}

Biaxial nematic phases attracting experimental, theoretical,
and computer simulation research since its first prediction by Freiser
\cite{1970_Freiser}.
There phases are characterized by an orientational order along
three perpendicular directions and by the existence of three
distinct optical axes. They are very interesting
from both the fundamental and the technological points of view
\cite{2008_Berardi_JP},
\cite{2010_Tschierske}.
Biaxial materials could offer a possibility of fast switching 
of the second director and better viewing characteristics.

In practical applications liquid crystals are always
placed in limited space and even a weak interaction
with a limiting surface can change the structure
of a liquid crystal near the boundary.
A basic model for a phase boundary is the smooth hard planar wall.
Despite its simplicity it can induce interesting phenomena.

The behavior of hard biaxial molecule fluids near a hard surface
is poorly understood.
In this paper we study the nematic-wall and the isotropic-wall
interfaces assuming that biaxial molecules interact
with one another and with the wall only via hard-core repulsion.
Analytical results for the surface tension and the entropy 
contributions are derived.
We find the preferred orientation 
of the phase composed of the most biaxial spheroplatelets
with the optimal shape between rods and plates.
For such molecules there is the direct transition from the isotropic phase
to the biaxial nematic phase on increasing the density.
The preferred phase orientation minimize the nematic-wall surface tension.
The density and order-parameter profiles are calculated 
in the case of the isotropic and the biaxial nematic phase.

The paper is organized as follows.
In Sec. \ref{sec:background} interfacial phenomena
and biaxial molecules studies are briefly reviewed
in order to provide the background for our studies.
In Sec. \ref{sec:theory} the statistical theory of the phase
ordering is provided for the case of the hard molecules
at the hard wall in the low-density limit.
In Sec. \ref{sec:results} the theory is applied to the system
of hard spheroplatelets where the direct transition
from the isotropic phase to the biaxial nematic phase is present in the bulk.
Section \ref{sec:summary} contains a summary.

\section{Background}
\label{sec:background}

In order to make the paper self-contained, we collect the relevant 
definitions and fact concerning interfacial phenomena
and biaxial molecule studies.

\subsection{Fluid interfacial phenomena}

There are many fluid interfacial phenomena, such as anchoring, 
critical adsorption, pre-wetting and wetting transitions
\cite{1991_Jerome}.
The possible structural rearrangements in the vicinity of the
interface are (1) periodic modulations of density,
(2) polar ordering of molecular dipoles,
and (3) modifications of the scalar order parameter
\cite{2003_Kleman_Lavrentovich}.
Anchoring is a fixing of the phase orientation by the surface
with lifting the orientation in the bulk via the elastic forces.
In confined geometry, phase transitions are usually shifted
with respect to the transitions observed in infinite geometry.

Let us consider the case of a second-order transition from
a disordered to an ordered phase. 
The order parameter fluctuations appear in the bulk with a correlation 
length which diverges at the transition.
The correlation length at the surface becomes infinite in a direction
parallel to the surface plane. This creates an ordered layer 
at the surface in which the order parameter decreases exponentially
to zero in the bulk over a penetration length.
The penetration length is equal to the correlation length
and thus diverges at the transition.
This phenomenon is called critical adsorption
\cite{1991_Jerome}.

When the transition is first order, the situation is more complex.
Partial or complete wetting can appear depending on the values
of a contact angle. When one explores the coexistence curve
between phases, one can go from a partial wetting regime to 
a complete wetting regime via a wetting transition.

\subsection{Studies of hard biaxial molecules}

Computer simulations studies of anisotropic hard molecules have 
confirmed that hard-core interactions are essential for liquid crystal 
phase behavior
\cite{1993_Allen}.
Over the years a variety of hard-particle models have been studied
theoretically and by using computer simulations.
These investigations have shown that hard-particle fluids
can exhibit many liquid-crystalline phases, such as
uniaxial and biaxial nematic
\cite{1990_Allen},
smectic, crystal, and plastic solid phases
\cite{1985_Frenkel_Mulder},
\cite{1987_Stroobants}.

Several types of biaxial molecule fluids were investigated:
ellipsoids with three different axes
\cite{1990_Allen},
\cite{1997_Camp_Allen},
\cite{1997_Vega},
\cite{1997_Zakhlevnykh_Sosnin},
\cite{2007_McBride_Lomba},
biaxial Gay-Berne particles
\cite{1995_Berardi},
\cite{1998_Berardi},
\cite{2000_Berardi_Zannoni},
\cite{2001_Zannoni},
\cite{2008_Berardi_JCP},
rectangular parallelepipeds
\cite{2008_John}, 
\cite{2011_Martinez-Raton},
spheroplatelets, and spherocuboids
\cite{2005_Mulder}.
Singh and Kumar developed a theory with a general convex-body 
coordinate system that can be used to describe any hard convex body
\cite{1996_Singh_Kumar},
\cite{2001_Singh_Kumar}.
The results can be utilized in the study of structural, thermodynamic,
and transport properties of ellipsoidal fluids.

The hard spheroplatelet is a natural generalization of the
spherocylinder. In 1986 Mulder expressed the pair-excluded volume
at fixed orientation in closed form
\cite{1986_Mulder}.
Later the phase diagram of the hard spheroplatelet fluid
was proposed as a result of bifurcation analysis
in the low-density Onsager approximation
\cite{1989_Mulder}.
The density versus particle biaxiality phase diagram
displays a cusp-shaped biaxial nematic phase intervening between
two uniaxial nematic phases.
Holyst and Poniewierski studied the Landau bicritical point
at which a direct transition from the isotropic phase 
to the biaxial nematic phase occurs
\cite{1990_Holyst_Poniewierski}.
A dense system of hard biaxial molecules (spheroplatelets
and ellipsoids) was considered using a density functional theory.
They found that the density of the isotropic phase at the Landau
bicritical point was always higher than that at the
isotropic-nematic transition in the limit of uniaxial molecules.

In 1991 Taylor extended the pair-excluded volume
to the case of non-identical spheroplatelets
\cite{1991_Taylor}.
In the same year Taylor and Herzfeld studied nematic and smectic
order in a fluid of hard biaxial spheroplatelets
\cite{1991_Taylor_Herzfeld}.
They used scaled particle theory for the fluid configurational
entropy, in conjunction with a cell description of
translational order. When the possibility of translational
order was considered, the phase diagram displayed three distinct
smectic A phases, columnar and crystalline ordering
for higher densities (packing greater then 0.6).
For low and intermediate densities the diagram was identical
with previous findings (the isotropic phase, the two uniaxial
nematic phases separated by the biaxial nematic phase).
The necessity of further studies of the Landau point region
was noted.

In 2009 van der Pol \emph{et al.} found biaxial nematic and biaxial 
smectic phases in a colloidal model system of mineral goethite particles with 
a simple boardlike shape and short-range repulsive interaction
\cite{2009_Pol}. 
The biaxial nematic phase was stable over a large concentration range
and the uniaxial nematic phase was not found.
Other studies showed that shape polidispersity of particles
can stabilize the biaxial nematic phase, and it can induce
a novel topology in the phase diagram
\cite{2003_Vanakaras},
\cite{2011_Belli}.
Another stabilizing factor is a small tetrahedral deformation
of particles as was shown within the extended Straley model
\cite{2011_Kapanowski}.

Recently Peroukidis \emph{et al.} calculated the full phase diagram
of hard  biaxial spheroplatelets by means of Monte Carlo simulations
\cite{2013_Peroukidis_Vanakaras},
\cite{2013_Peroukidis}.
New classes of phase sequences were identified:
$I-[N_{+}]-SmA$, 
$I-[N_{b-}-N_{b+}]-SmA$ (crossover),
$I-[N_{-}]-SmA$, 
$I-[N_{-}]-Col_x$ (columnar phases),
$I-Cub$ (cubatic phases).
The brackets indicate phases that may be absent.
The most interesting finding was the crossover between two
distinct biaxial nematic states.
The formation of anisotropic supramolecular assemblies was demonstrated.

\subsection{Hard molecules at the interface}

Properties of liquid crystal phases in the bulk and at the surface
generally are not the same. Different physical systems were studied 
in the past: fluids with uniaxial molecules in contact with a single 
(hard or attractive) wall, confined by two walls (thin cells)
or curved surfaces
\cite{2012_Zhang}.
Let us recall the main results concerning solid-fluid interfaces.
We will not discuss nematic free surfaces and thin films.

In 1984 Telo da Gama studied wetting transitions at a solid-fluid interface
using attractive walls and the attractive forces
with a hard core for molecular interactions
\cite{1984_Telo_da_Gama}.
The wetting transitions were always weakly first order.
In 1988 Poniewierski and Holyst studied a system of hard spherocylinders
in contact with a single hard wall
\cite{1988_Poniewierski_Holyst_PRA}.
They used a simple local approximation for the one-particle 
distribution function and showed that the preferred orientation 
of the nematic director is parallel to the wall. 
The density and order-parameter profiles were calculated. 
The nematic main order parameter was enhanced near the wall even 
though the density was reduced.
The wall-induced biaxiality was small in the interfacial region.
Wetting by the nematic phase occurred at the
nematic-isotropic coexistence.
Later the stability of the uniaxial solution close to the wall
was investigated in the limit of very long molecules
\cite{1993_Poniewierski},
and the bifurcation point was found.
The nematic-phase--isotropic-phase interface for hard spherocylinders
was studied in Ref.
\cite{1988_Holyst_Poniewierski_PRA}.

A hard-rod fluid confined by two parallel wall 
was studied by Mao \emph{et al.}
\cite{1997_Mao}.
The aim of this work was to calculate the depletion 
force between the plates due to confinement of the rods.
Van Roij \emph{et al.} investigated the phase behavior of colloidal 
hard-rod fluids ($L/D=15$)
near a single wall and confined in a slit pore
\cite{2000_Roij},
\cite{2000_Roij_JCP},
\cite{2001_Dijkstra},
\cite{2005_Dijkstra}.
They obtained (1) a wall-induced surface transition from uniaxial
to biaxial symmetry, (2) complete orientational wetting of the
wall-isotropic fluid interface by a nematic film, and (3)
capillary nematization, with a capillary critical point,
induced by confinement in the slit pore.

The properties of a model suspension of hard colloidal platelets
with continuous orientations and vanishing thickness were studied
using several methods by Reich \emph{et al.}
\cite{2007_Reich}.
It is interesting that this system is not described well by the Onsager
theory, and a scaling argument known from thin rods does not hold.
The fundamental measure theory density functional was used,
which includes contributions to the free energy that are of the third
order in density.

\section{Theory}
\label{sec:theory}

The aim of this section is to develop the statistical theory of the phase
ordering for the case of the hard molecules
at the hard wall in the low-density limit.
The expressions for the density, the order parameters, and 
the surface tension will be derived.

\subsection{Description of the system}

\begin{figure}
\begin{center}
\includegraphics[width=0.5\textwidth]{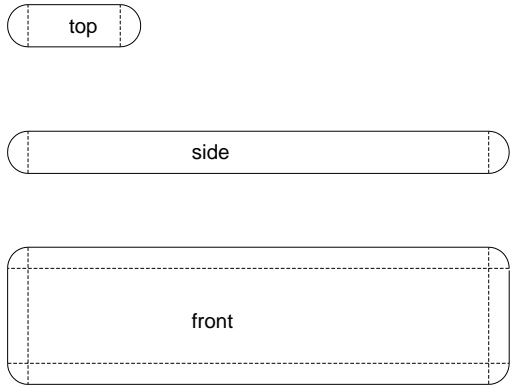}
\end{center}
\caption[Spheroplatelet.]{
\label{fig1}
Front, side, and top view of a spheroplatelet.}
\end{figure}

The system of hard spheroplatelets in the presence
of a hard wall is considered. 
A spheroplatelet can be described as a rectangular
block with dimensions $2a \times b \times c$, capped with quarter
spheres of radius $a$ and half-cylinders with radius $a$
and lengths $b$ and $c$ such as to produce a piece-wise
smooth convex body; see Fig.~\ref{fig1}.
The position and the orientation of a spheroplatelet are determined
by $\vec{r}$ and the three Euler angles $R=(\phi,\theta,\psi)$, 
respectively.
Alternatively, the orientation can be described by the three
orthonormal vectors $(\vec{l},\vec{m},\vec{n})$.
The $z$ axis is chosen to be perpendicular to the wall.
The density of the fluid at $z=+\infty$ is $\rho_0$.

The grand thermodynamical potential $\Omega$ 
as a functional of the one-particle distribution
function $\rho (\vec{r},R)$ has the following form:
\begin{multline}
\label{eq:beta_omega}
\beta \Omega \{\rho\} = \beta F_{id} \{ \rho \}
+ \beta F_{ex} \{ \rho \}   \\
+ \beta \int d\vec{r} dR \rho (\vec{r},R) [V_{ext} (\vec{r},R)-\mu],
\end{multline}
where the ideal gas contribution is

\begin{equation}
\beta F_{id} \{ \rho \} = \int d\vec{r} dR \rho (\vec{r},R)
\{ \ln [ \Lambda \rho (\vec{r},R) ] -1 \},
\end{equation}
$\mu$ is the chemical potential,
$\beta = 1/k_B T$ is the Boltzmann factor,
$V_{ext}$ stands for the external potential,
and $\Lambda$ is the (irrelevant) thermal volume of molecules.
$F_{ex}$ is the excess part of the free energy corresponding 
to the interactions between molecules.
We assume the low-density Onsager approximation 
for $F_{ex}$, i.e.,

\begin{equation}
\beta F_{ex} = - \frac{1}{2} \int d\vec{r}_1 dR_1 d\vec{r}_2 dR_2
\rho (\vec{r}_1,R_1) \rho (\vec{r}_2,R_2) f_{12},
\end{equation}
where $f_{12}$ stands for the Mayer function, which is equal
to $-1$ when two molecules overlap and $0$ otherwise.
The one-particle distribution function has the normalization

\begin{equation}
\int  d\vec{r} dR \rho (\vec{r},R) = N.
\end{equation}
The expression for the external potential exerted on a molecule
by the hard wall reads as follows:

\begin{equation}
V_{ext} (z,R) = 
\left\{
\begin{array}{ll}
+\infty   & \mbox{for} \ z < z_m(R),  \\
0         & \mbox{for} \ z > z_m(R),
\end{array}
\right.
\end{equation}
where $z_m(R) = a+(b|m_z|+c|n_z|)/2$ stands for the minimal
distance between the wall and a molecule of orientation $R$.
The minimization of $\Omega \{\rho\}$ with respect to 
$\rho (\vec{r},R)$ leads to the integral equation for $\rho (\vec{r},R)$:

\begin{multline}
\label{eq:ln_ro}
\ln [\Lambda \rho (\vec{r}_1,R_1)] 
+ \beta V_{ext} (\vec{r}_1,R_1)      \\
- \int d\vec{r}_2 dR_2 \rho (\vec{r}_2,R_2) f_{12} = \beta \mu.
\end{multline}
In the absence of an external potential Eq. (\ref{eq:ln_ro})
has a spatially uniform solution
$\rho (\vec{r}_1,R_1) = \rho_0 f(R)$, where $f(R)$
is the orientational distribution function normalized to unity.
For the isotropic phase $f(R)=1/8\pi^2$,
for the uniaxial nematic phase 
$f(R)=f(\vec{l} \cdot \vec{N}, \vec{n} \cdot \vec{N})$,
and for the biaxial nematic phase
$f(R)=f(\vec{l} \cdot \vec{L}, \vec{l} \cdot \vec{N},
\vec{n} \cdot \vec{L}, \vec{n} \cdot \vec{N})$.
The unit orthogonal vectors $(\vec{L},\vec{M},\vec{N})$
determine three axes of the $D_{2h}$ symmetry of the biaxial
nematic phase. In the uniaxial nematic phase with $D_{\infty h}$
symmetry only the $\vec{N}$ vector survives.

\subsection{The liquid crystal-wall surface tension}

When the wall is present, instead of solving Eq. (\ref{eq:ln_ro}),
we approximate $\rho (z,R)$ as follows
\cite{1988_Poniewierski_Holyst_PRA}:

\begin{equation}
\label{eq:ro_approx}
\rho (z,R) = \rho_0 f(R) \exp [-\beta V_{ext} (z,R)].
\end{equation}
Let us note that the density profiles obtained from (\ref{eq:ro_approx})
will not exhibit the short-range oscillatory behavior
that is expected close to the wall.
It is assumed that the directors $(\vec{L},\vec{M},\vec{N})$
do not change throughout the sample.
Substitution of (\ref{eq:ro_approx}) into (\ref{eq:beta_omega})
and subtraction of the bulk term leads to the expression
for the liquid crystal-wall surface tension $\gamma$
\cite{1988_Poniewierski_Holyst_PRA},

\begin{equation}
\label{eq:tension}
\beta \gamma = -(S_{rot}+S_{tr,id}+S_{tr,ex})/k_B -\beta \Delta \mu \Gamma,
\end{equation}
where $S_{rot}$, $S_{tr,id}$, and $S_{tr,ex}$ are
the surface entropies per unit area,

\begin{equation}
\beta \Delta \mu = \beta \mu - \ln [\Lambda \rho_0 /(8\pi^2)];
\end{equation}

\begin{equation}
\begin{split}
\Gamma &= \int_{0}^{\infty} dz dR [\rho(z,R)-\rho_0 f(R)]  \\
&= -\rho_0 \int dR f(R) z_m (R)
\end{split}
\end{equation}
stands for the adsorption
\cite{1988_Poniewierski_Holyst_PRA}.
The rotational entropy $S_{rot}$ comes only from the ideal term
in the free energy. According to the usual convention,
the rotational entropy is defined in such
a way that it vanishes for the isotropic phase

\begin{equation}
S_{rot}/k_B = \rho_0 \int dR f(R) z_m (R) \ln [8 \pi^2 f(R)].
\end{equation}
The translational entropy have two contributions:
$S_{tr,id}$ from the ideal term and
$S_{tr,ex}$ from the excess term:

\begin{equation}
S_{tr,id}/k_B = -\rho_0 \int dR f(R) z_m (R),
\end{equation}


\begin{multline}
S_{tr,ex}/k_B  =  \\
\frac{1}{2} \rho_0^2 
\int dR_1 dR_2 f(R_1) f(R_2) z_m(R_1) K(R_1, R_2)  \\
+ \frac{1}{2} \rho_0^2 
\int dR_1 dR_2 f(R_1) f(R_2) L(R_1, R_2),
\end{multline}
where
\begin{multline}
L(R_1,R_2) = \int_{z_m(R_1)-z_m(R_2)}^{\infty} dz_{12}  \\
\times \left[ z_{12}-z_m(R_1)+z_m(R_2) \right]
V(|z_{12}|,R_1,R_2),
\end{multline}


\begin{equation}
V(|z_{12}|,R_1,R_2) = -\int dx_{12} dy_{12} f_{12},
\end{equation}

\begin{equation}
K(R_1, R_2) = -\int dr_{12} f_{12}.
\end{equation}
$K(R_1, R_2)$ is the excluded volume for two spheroplatelets.
$V(|z_{12}|,R_1,R_2)$ is the intersection of the excluded volume
for two spheroplatelets of orientations $R_1$ and $R_2$
with a plane parallel to the wall and distant from the center
of the excluded volume by $|z_{12}|$.
The entropy $S_{tr,id}$ is negative because the wall restricts
the translational freedom of molecules.
The first (positive) term in $S_{tr,ex}$ takes into account
the pairs of molecules, one of which interacts directly with
the wall whereas the other does not.
The second (positive) term takes into account all pairs
in which both molecules interact directly with the wall
\cite{1988_Poniewierski_Holyst_PRA}.

The nematic-wall surface tension $\gamma$
is a function of directors through the distribution function $f(R)$.
The tension should be minimized with respect to the phase orientation
in order to find the equilibrium value of the phase orientation.

\subsection{The density and the order parameter profiles}

In the approximation (\ref{eq:ro_approx}) for the one-particle
distribution function $\rho (z,R)$, the thickness of the interfacial
region is equal to the range of $V_{ext}$.
Outside, the density and the order parameters are equal
to their bulk values. Thus only the range
$ a \le z \le a + \sqrt{b^2 + c^2}/2 $ is interesting.
For $z < a$, $\rho (z) = 0$ and the order parameters are undefined.
Integrating $\rho (z,R)$ over the angular variables, we find that

\begin{equation}
\rho (z) = \rho_0 \int dR f(R) \exp [-\beta V_{ext} (z,R)].
\end{equation}
The orientational distribution function is equal to
$f(z,R) = \rho(z,R)/\rho(z)$
in the interfacial region.
The average of any function $A(R)$ can be calculated as

\begin{equation}
\label{eq:mean_A}
\langle A \rangle (z) = \int dR f(z,R) A(R).
\end{equation}

The formula (\ref{eq:mean_A}) will be used to calculate
the order parameters.

\section{Results}
\label{sec:results}

The spheroplatelets are useful objects because many calculations
can be done analytically. In this section the most important
results from the literature are recalled and an exemplary calculations
for the spheroplatelets at the hard wall are presented.

\subsection{Spheroplatelets}

The volume of a spheroplatelet is equal to

\begin{equation}
V_{mol} = 4 \pi a^3/3 + \pi a^2 (b+c) + 2abc.
\end{equation}
The pair-excluded volume is given by
\cite{1989_Mulder}

\begin{multline}
K (R_1, R_2)  =  32 \pi a^3/3 + 8\pi a^2 (b+c) + 8abc    \\
+ 4abc \{ |\vec{m}_1 \times \vec{n}_2| 
+ |\vec{n}_1 \times \vec{m}_2| \}    \\
+ 4ab^2 |\vec{m}_1 \times \vec{m}_2| 
+ 4ac^2 |\vec{n}_1 \times \vec{n}_2|      \\
+ b^2 c \{ |\vec{l}_1 \cdot \vec{m}_2| 
+ |\vec{m}_1 \cdot \vec{l}_2| \}    \\
+ bc^2 \{ |\vec{l}_1 \cdot \vec{n}_2| 
+ |\vec{n}_1 \cdot \vec{l}_2| \} .
\label{excluded_mulder}
\end{multline}
The expansion of the excluded volume $K(R_1, R_2)$ can be given as
\begin{equation}
\begin{split}
K(R_1, R_2) &= \sum_j \sum_{\mu\nu} K_{\mu\nu}^{(j)}
F_{\mu\nu}^{(j)} (R_2^{-1} R_1)  \\
&= \sum_{[I]} K^{[I]} F^{[I]} (R_2^{-1} R_1),
\end{split}
\end{equation}
where the coefficients $K_{\mu\nu}^{(j)}$ are symmetric
in the indices $\mu$ and $\nu$ due to the particle
interchange symmetry.
The invariants $F_{\mu\nu}^{(j)}=F^{[I]}$ are defined in Ref.
\cite{1997_Kapanowski}, where the indicator $[I]=(j,\mu,\nu)$
is explained. First indices are:
$[1]=(0,0,0)$, $[2]=(2,0,0)$, $[3]=(2,0,2)$,
$[4]=(2,2,0)$, and $[5]=(2,2,2)$.
The invariants are related to Wigner functions
$D_{\mu\nu}^{(j)}$. If $j$ is even, then $0 \le \mu,\nu \le j$,

\begin{equation}
F_{00}^{(j)}(R) = D_{00}^{(j)}(R),
\end{equation}

\begin{equation}
F_{0\nu}^{(j)}(R) = \frac{1}{\sqrt{2}} 
[D_{0\nu}^{(j)}(R) + D_{0-\nu}^{(j)}(R)],
\end{equation}

\begin{equation}
F_{\mu 0}^{(j)}(R) = \frac{1}{\sqrt{2}} 
[D_{\mu 0}^{(j)}(R) + D_{-\mu 0}^{(j)}(R)],
\end{equation}

\begin{equation}
F_{\mu\nu}^{(j)}(R) = \frac{1}{2} 
[D_{\mu\nu}^{(j)}(R) + D_{\mu -\nu}^{(j)}(R)
+ D_{-\mu\nu}^{(j)}(R) + D_{-\mu -\nu}^{(j)}(R)].
\end{equation}
If $j$ is odd, then $2 \le \mu,\nu \le j$,
\begin{equation}
F_{\mu\nu}^{(j)}(R) = \frac{1}{2} 
[D_{\mu\nu}^{(j)}(R) - D_{\mu -\nu}^{(j)}(R)
- D_{-\mu\nu}^{(j)}(R) + D_{-\mu -\nu}^{(j)}(R)].
\end{equation}
The most important excluded volume coefficients have the form:

\begin{multline}
K_{00}^{(0)} = 32\pi a^3/3 + 8\pi a^2(b+c) + (8+2\pi)abc   \\
+ \pi a(b^2+c^2) + b^2 c + bc^2,
\end{multline}

\begin{equation}
K_{00}^{(2)} = (5/16)(b^2 c -2bc^2 + 2\pi abc -2\pi ac^2 -\pi ab^2/2),
\end{equation}

\begin{equation}
K_{02}^{(2)} = K_{20}^{(2)} 
= (5\sqrt{3}/16)(bc^2 + \pi abc -\pi ab^2/2),
\end{equation}

\begin{equation}
K_{22}^{(2)} = (-15/16)(b^2 c + \pi a b^2/2).
\end{equation}
For molecules intermediate between rods and plates,
called the most biaxial molecules,
the direct transition from the isotropic to the biaxial
nematic phase is present.
In that case $K_{02}^{(2)} = K_{20}^{(2)} = 0$, 
$K_{00}^{(2)} > 0$, $K_{22}^{(2)} < 0$, and

\begin{equation}
c^2 + \pi ac -\pi ab/2 = 0 \ \mbox{for} \ b>c.
\end{equation}
From the analysis of isotropic-symmetry-breaking bifurcations 
\cite{1989_Mulder}
it is possible to find the transition point from the isotropic phase
to the biaxial nematic phase

\begin{equation}
\rho_C = -5/K_{22}^{(2)}\ \mbox{for}\ b>c.
\end{equation}
We will study physically equivalent systems with $b<c$
because then it is easier to discuss the values
of the order parameters.
The condition for the most biaxial molecules has the form

\begin{equation}
b^2 + \pi ab -\pi ac/2 = 0 \ \mbox{for} \ b<c.
\end{equation}
Let us define packing $y = \rho V_{mol}$.
We studied two systems with $b<c$ in order to check that
our results do not depend qualitatively on the molecule elongations
(this is important in the context of the Onsager approximation):

\begin{gather}
b=1.5\pi a,\ c=7.5\pi a,\ y_C = 0.48174\ \mbox{(system A)}, \\
b=2\pi a,\ c=12\pi a,\ y_C = 0.29437\ \mbox{(system B)}.
\end{gather}
This choice corresponds to the following sets of parameters
from Ref.
\cite{2013_Peroukidis_Vanakaras}:
$(l^*, w^*) \approx (12.78, 3.36)$ for system A;
$(l^*, w^*) \approx (19.85, 4.14)$ for system B.

\subsection{The phase in the bulk}

\begin{figure}
\begin{center}
\includegraphics[width=0.5\textwidth]{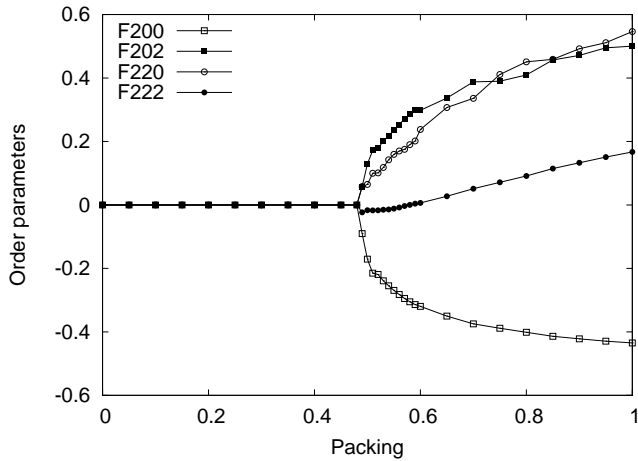}
\end{center}
\caption[Order parameters vs packing in the bulk (system A).]{
\label{fig2}
Order parameters $\langle F_{\mu\nu}^{(2)} \rangle$
($F2\mu\nu$ in the picture)
vs packing in the bulk for system A.
The phase orientation for $y>y_C$ is described by the vectors
$(\vec{L},\vec{M},\vec{N})=(\vec{e}_z,-\vec{e}_y,\vec{e}_x)$.
This is the solution minimizing the surface tension.
}
\end{figure}

\begin{figure}
\begin{center}
\includegraphics[width=0.5\textwidth]{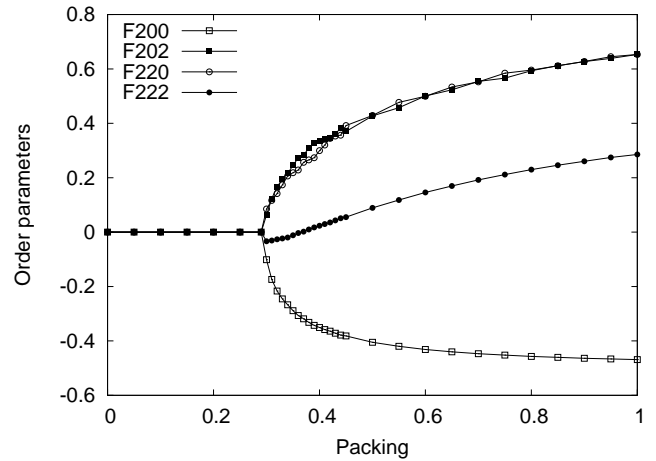}
\end{center}
\caption[Order parameters vs packing in the bulk (system B).]{
\label{fig3}
Order parameters $\langle F_{\mu\nu}^{(2)} \rangle$
($F2\mu\nu$ in the picture) vs packing in the bulk for system B.
The phase orientation is the same as in Fig.~\ref{fig2}.
}
\end{figure}

The phase in the bulk is described by a spatially uniform solution
of the form

\begin{equation}
\ln f(R) = \sum_{j} \sum_{\mu\nu} S_{\mu\nu}^{(j)} F_{\mu\nu}^{(j)}(R),
\end{equation}

\begin{equation}
\label{eq:Sjmn}
S_{\mu\nu}^{(j)} = -\rho_0 \sum_{\sigma} K_{\sigma\nu}^{(j)}
\langle F_{\mu\sigma}^{(j)} \rangle \ \mbox{for} \ j>0,
\end{equation}

\begin{equation}
\langle F_{00}^{(0)} \rangle = 1 \ \mbox{(the normalization condition)}.
\end{equation}
According to Mulder 
\cite{1989_Mulder}
and others
\cite{2006_Rosso_Virga},
we can focus on the $j=2$ subspace with four independent
parameters $S_{\mu\nu}^{(2)}$,

\begin{equation}
f \sim \exp(S^{[2]} F^{[2]} 
+ S^{[3]} F^{[3]} 
+ S^{[4]} F^{[4]} 
+ S^{[5]} F^{[5]}).
\end{equation}
The solution of Eq.~(\ref{eq:Sjmn}) should have the orientation
minimizing the surface tension $\gamma$.
Equation (\ref{eq:Sjmn}) was solved numerically
for systems A and B by means of the C program.
Multidimensional minimization was done by the downhill simplex method,
implemented in the function \emph{amoeba}
\cite{Numerical_Recipes}.
The dependence of the order parameters on the packing
are presented in Fig.~\ref{fig2} (system A) and
in Fig.~\ref{fig3} (system B).

Let us recall the meaning of the order parameters 
$\langle F_{\mu\nu}^{(2)} \rangle$.
The $\langle F_{00}^{(2)} \rangle$ order parameter is a measure
of the alignment of the $\vec{n}$ molecule axis along the $Z$ axis
of the reference frame.
The $\langle F_{02}^{(2)} \rangle$ order parameter
describes the relative distribution of the $\vec{l}$
and the $\vec{m}$ axes along the $Z$ axis.
Both $\langle F_{00}^{(2)} \rangle$ and $\langle F_{02}^{(2)} \rangle$
can be nonzero in the uniaxial nematic phase.
The $\langle F_{20}^{(2)} \rangle$ order parameter
describes the relative distribution of the $\vec{n}$ axis
along the $X$ and the $Y$ axes.
The $\langle F_{22}^{(2)} \rangle$ order parameter
is related to the distribution
of the $\vec{l}$ axis along the $X$ axis and the distribution
of the $\vec{m}$ axis along the $Y$ axis.

\subsection{Isotropic phase in the bulk}

\begin{figure}
\begin{center}
\includegraphics[width=0.5\textwidth]{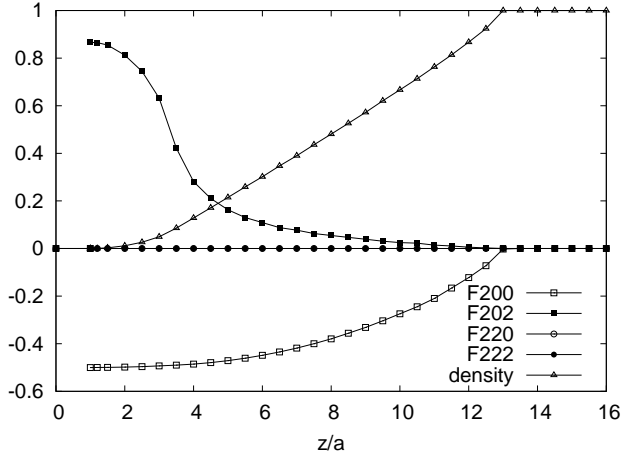}
\end{center}
\caption[System A in the isotropic phase.]{
\label{fig4}
Density profile $\rho(z)/\rho_0$ and order parameters 
$\langle F_{\mu\nu}^{(2)} \rangle$ ($F2\mu\nu$ in the picture)
for the isotropic phase in the bulk (system A).
In the interfacial region the phase is uniaxial,
$\langle F_{20}^{(2)} \rangle = \langle F_{22}^{(2)} \rangle = 0$.
The density decreases to zero near the wall.
The molecule positions are in the range $z>a$.}
\end{figure}

The system is in the isotropic phase ($S_{\mu\nu}^{(j)}=0$)
for $\rho < \rho_C$.
Near the wall the order parameter $\langle F_{00}^{(2)} \rangle$
decreases from 0 to $-1/2$ whereas the order parameter
$\langle F_{02}^{(2)} \rangle$ increases from 0 to $\sqrt{3}/2$.
The long molecule axes $\vec{n}$ tend to be parallel to the wall and 
the short molecule axes $\vec{l}$ tend to be perpendicular to the wall.
The symmetry in the $xy$ plane is not broken and the phase is uniaxial.
In the interfacial region the density is reduced and decreases
to zero as $z \rightarrow a$.
The density and the order parameters profiles for system A
are plotted in Fig.~\ref{fig4}.
In the case of system B the results are similar.
The same profiles for the uniaxial order parameter 
$\langle F_{00}^{(2)} \rangle$
were obtained in the case of hard spherocylinders
by means of Monte Carlo simulations
\cite{2001_Dijkstra}. 
We have the additional nonzero order parameter
$\langle F_{02}^{(2)} \rangle$ indicating that
our particles are biaxial.
The density profiles of a hard-spherocylinder fluid suggest that
a small kink (a density maximum) at $z=a+c/2$ is possible
for the spheroplatelets.

The surface tension for the isotropic phase is positive
and has the form
\begin{multline}
\beta \gamma = \rho_0 (a+b/4+c/4) \\
+ \frac{1}{2} \rho_0^2 [K^{[1]} (a+b/4+c/4) - L^{[1][1]}].
\end{multline}
The relation between $\mu$ and $\rho_0$ in our model, 
for the case of the isotropic phase and the weak biaxial nematic 
phase, is
\begin{equation}
\beta \Delta \mu = \rho_0 K^{[1]} > 0.
\end{equation}

\subsection{Weak biaxial nematic  phase}

Let us consider a weak biaxial nematic phase near the transition point
from the isotropic to the biaxial nematic phase:

\begin{equation}
S_{\mu\nu}^{(j)} \ll 1 \ \mbox{for} \ j>0,\ S_{00}^{(0)} = - \ln (8 \pi^2).
\end{equation}
The orientational distribution function $f(R)$ and the order
parameters have a simplified form

\begin{equation}
f(R) = \frac{1}{8 \pi^2} \left[
1 + \sum_{j>0} S_{\mu\nu}^{(j)} F_{\mu\nu}^{(j)}(R)
\right],
\end{equation}

\begin{equation}
\label{eq:Fjmn_weak}
\langle F_{\mu\nu}^{(j)} \rangle = \frac{S_{\mu\nu}^{(j)}}{2j+1}.
\end{equation}
For the case of the weak biaxial nematic phase, it is possible
to calculate many physical quantities analytically.

\subsection{Alignment close to the wall}

Disregarding the problem of the equilibrium phase orientation
we can study the alignment close to the wall $(z=a)$
in the case of the weak biaxial nematic phase in the bulk.
The restrictions imposed on the Euler angles are as follows:
$\theta = \pi/2$, 
$\psi = 0$ or $\psi = \pi$, or $\psi = 2\pi$.
We assume that the most important are the parameters
$S_{\mu\nu}^{(2)}$ and the order parameters
$\langle F_{\mu\nu}^{(2)} \rangle$:

\begin{gather}
F_{00}^{(2)} (R) |_{wall} = -1/2, \\
F_{02}^{(2)} (R) |_{wall} = \sqrt{3}/2,  \\
F_{20}^{(2)} (R) |_{wall} = \cos(2 \phi) \sqrt{3}/2,  \\
F_{22}^{(2)} (R) |_{wall} = \cos(2 \phi)/2.
\end{gather}

\begin{equation}
f(a, \phi) = \frac{1}{2 \pi} \left[
1 + \sum_{\mu\nu} S_{\mu\nu}^{(2)} F_{\mu\nu}^{(2)}(R) |_{wall}
\right],
\end{equation}

\begin{equation}
\langle F_{\mu\nu}^{(2)} \rangle |_{wall}
= \int_{0}^{2 \pi} d\phi f(a,\phi) F_{\mu\nu}^{(2)}(R) |_{wall},
\end{equation}

\begin{equation}
\langle F_{00}^{(2)} \rangle |_{wall} = -1/2,
\end{equation}

\begin{equation}
\langle F_{02}^{(2)} \rangle |_{wall} = \sqrt{3}/2,
\end{equation}

\begin{equation}
\langle F_{20}^{(2)} \rangle |_{wall} =
\left[ S_{20}^{(2)} \sqrt{3} + S_{22}^{(2)} \right] \sqrt{3} \pi/4,
\end{equation}

\begin{equation}
\langle F_{22}^{(2)} \rangle |_{wall} =
\left[ S_{20}^{(2)} \sqrt{3} + S_{22}^{(2)} \right] \pi/4.
\end{equation}
The order parameters close to the wall can be expressed
by the bulk order parameters by means of Eq. (\ref{eq:Fjmn_weak}):

\begin{equation}
\langle F_{20}^{(2)} \rangle |_{wall} =
\left[ \langle F_{20}^{(2)} \rangle \sqrt{3}
+ \langle F_{22}^{(2)} \rangle \right] 5 \sqrt{3} \pi/4,
\end{equation}

\begin{equation}
\langle F_{22}^{(2)} \rangle |_{wall} =
\left[ \langle F_{20}^{(2)} \rangle  \sqrt{3}
+ \langle F_{22}^{(2)} \rangle \right] 5\pi/4.
\end{equation}
We conclude that biaxiality close to the wall can appear
only if the phase is biaxial in the bulk.
Note that the equality
$\langle F_{20}^{(2)} \rangle |_{wall} =
\sqrt{3} \langle F_{22}^{(2)} \rangle |_{wall}$
is valid also for the strong biaxial nematic phase.

\subsection{Alignment in the interfacial region}

The weak biaxial phase is now considered.
It is possible to calculate almost all parts of the surface
tension analytically:

\begin{multline}
\Gamma = -\rho_0 [ ( a + b/4 + c/4 ) \\
+ S^{[2]} ( -b/32 + c/16 )  
+ S^{[3]} ( -\sqrt{3} b/32 ) ],
\end{multline}

\begin{multline}
S_{rot}/(\rho_0 k_B) = 
S^{[2]} ( -b/32 + c/16 ) + S^{[3]} ( -\sqrt{3} b/32 )  \\
+ S^{[2]} S^{[2]} ( a/5 + 5b/128 + c/16 )  \\
+ S^{[2]} S^{[3]} ( \sqrt{3} b/64  )   \\
+ S^{[3]} S^{[3]} ( a/5 + 7b/128 + c/32 )  \\
+ S^{[4]} S^{[4]} (a/5 + 15b/256 + c/32) \\
+ S^{[4]} S^{[5]} ( -7\sqrt{3} b/384 ) \\
+ S^{[5]} S^{[5]} (a/5 + 31b/768 + 13c/24),
\end{multline}

\begin{equation}
S_{tr,id}/k_B = \Gamma,
\end{equation}

\begin{equation}
A_{\mu\sigma}^{(j)} = \frac{1}{2j+1} \sum_{\nu} 
S_{\mu\nu}^{(j)} K_{\sigma\nu}^{(j)},
\end{equation}

\begin{equation}
L^{[I][J]} = \int dR_1 dR_2 \frac{1}{(8\pi^2)^2}
F^{[I]}(R_1) F^{[J]}(R_2) L(R_1, R_2),
\end{equation}

\begin{multline}
2 S_{tr,ex}/(\rho_0^2 k_B) = K^{[1]} (a+b/4+c/4)  \\
+ A^{[2]} (-b/32+c/16) + A^{[3]} (-b \sqrt{3}/32)  \\
+ K^{[1]} S^{[2]} (-b/32+c/16) + K^{[1]} S^{[3]} (-b \sqrt{3}/32)  \\
+ S^{[2]} A^{[2]} ( a/5 + 5b/128 + c/16 )  \\
+ (S^{[2]} A^{[3]} + S^{[3]} A^{[2]}) (b \sqrt{3}/128)  \\
+ S^{[3]} A^{[3]} (a/5 + 7 b/128 + c/32)   \\
+ S^{[4]} A^{[4]} (a/5 + 15 b/256+ c/32)   \\
+ (S^{[4]} A^{[5]} + S^{[5]} A^{[4]}) (-7 \sqrt{3} b/768)   \\
+ S^{[5]} A^{[5]} (a/5 + 31 b/768 + 13 c/24)  \\
+ L^{[1][1]} + \sum_{[I]=2}^{5} S^{[I]} (L^{[1][I]} + L^{[I][1]})  \\
+ \sum_{[I]=2}^{5} \sum_{[J]=2}^{5} S^{[I]} S^{[J]} L^{[I][J]}. 
\end{multline}
The coefficients $L^{[I][J]}$ were calculated numerically in two steps.
In the first step, the values of the function $L(R_1,R_2)$ were calculated
for the selected orientations $(R_1, R_2)$ by means of Romberg's method
\cite{Numerical_Recipes}. 
The function $V(|z_{12}|,R_1,R_2)$ was calculated in the discrete space 
where the space step length was $a/2$ or $a/3$.
In the second step, the Gauss-Legendre integration in six dimensions
(six Euler angles) was applied.
The approximations with four, eight, and 16 nodes per dimension
were checked.
The programs were implemented in Python and C++ languages.
In Tables \ref{tab1} and \ref{tab2} the coefficients $L^{[I][J]}$
are reported, obtained with 16 nodes per dimension.
Errors estimated were less then 10\%.

Let us note that in the case of hard ellipsoids the hard Gaussian
overlap (HGO) model 
\cite{1972_Berne_Pechukas},
\cite{1998_Berardi}
is often used, because it is computationally simple
and shares some similarities with the hard ellipsoid fluid.
However, it was shown
\cite{2001_Miguel_Rio}
that the HGO model turns out to be inappropriate for elongated
molecules (length to breadth ratio above 5).
In the case of spheroplatelets we used known expressions for the
excluded volume and the $K^{[I]}$ coefficients, 
but the coefficients $L^{[I][J]}$ were calculated numerically.
Inside the formula for the surface tension there are no terms
with $K^{[I]}$ that mix $(S^{[2]}, S^{[3]})$ with $(S^{[4]}, S^{[5]})$.
Numerical calculations suggest that the same is true for
the terms with $L^{[I][J]}$.
The biaxial order parameters are separated from the uniaxial ones.

\begin{figure}
\begin{center}
\includegraphics[width=0.5\textwidth]{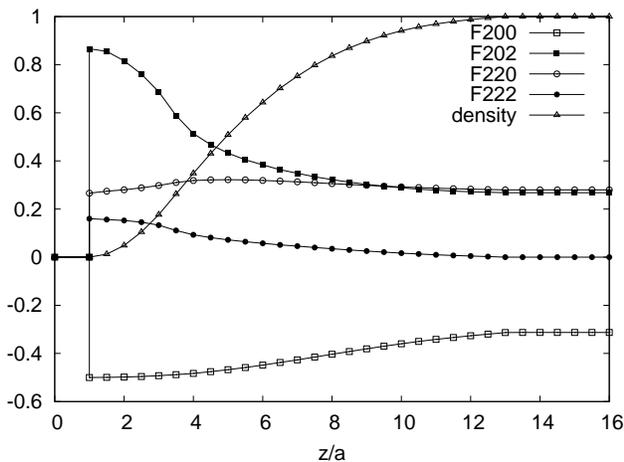}
\end{center}
\caption[System A in the biaxial nematic phase.]{
\label{fig5}
Density profile $\rho(z)/\rho_0$ and order parameters 
$\langle F_{\mu\nu}^{(2)} \rangle$ ($F2\mu\nu$ in the picture)
for the biaxial nematic phase in the bulk at packing $y=0.6$
(system A). In the interfacial region the phase is biaxial.
The density decreases to zero near the wall.
The phase orientation is
$(\vec{L},\vec{M},\vec{N})=(\vec{e}_z,-\vec{e}_y,\vec{e}_x)$.}
\end{figure}

\begin{figure}
\begin{center}
\includegraphics[width=0.5\textwidth]{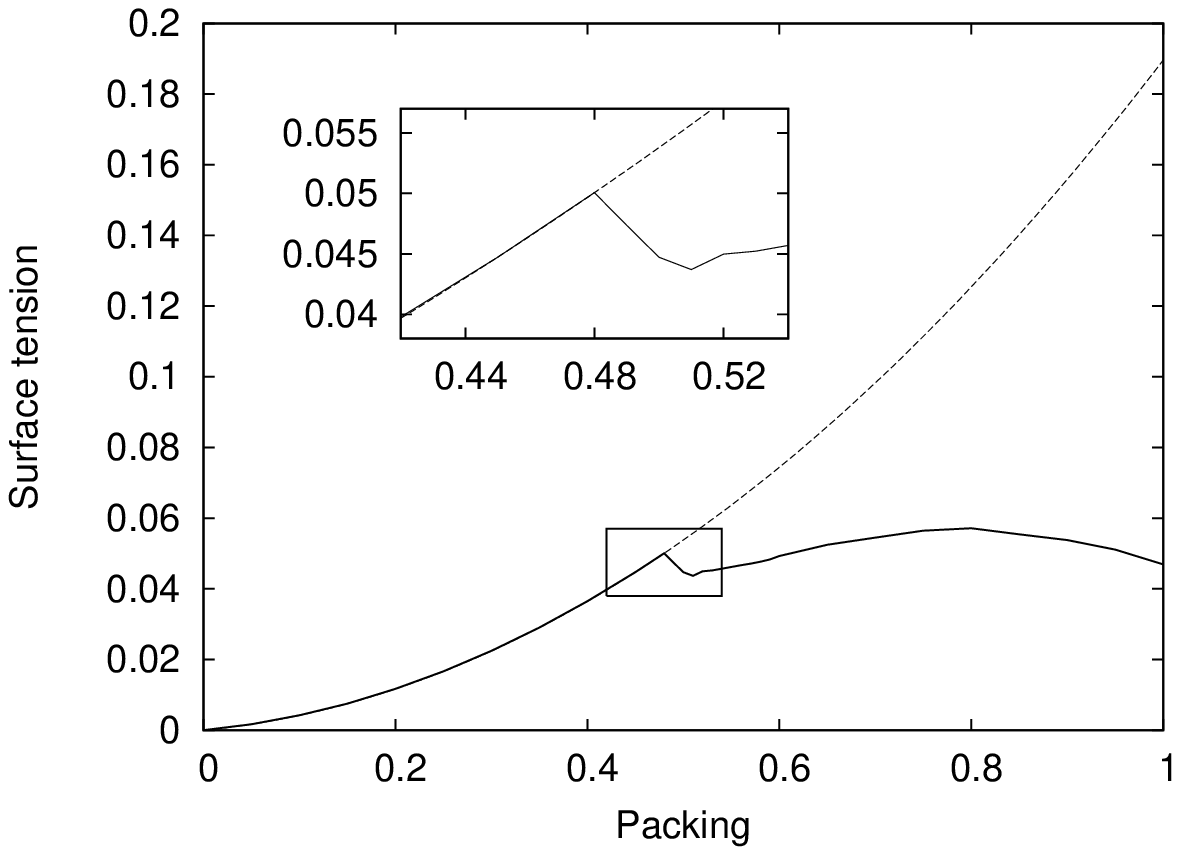}
\end{center}
\caption[Surface tension vs packing for system A).]{
\label{fig6}
Surface tension $\beta \gamma a^2$ vs packing (system A).
The dashed line describes values calculated for the isotropic phase.
The inset shows the neighborhood of the point $y_C=0.48174$
with the transition from the isotropic phase to the biaxial nematic phase.}
\end{figure}

\begin{figure}
\begin{center}
\includegraphics[width=0.5\textwidth]{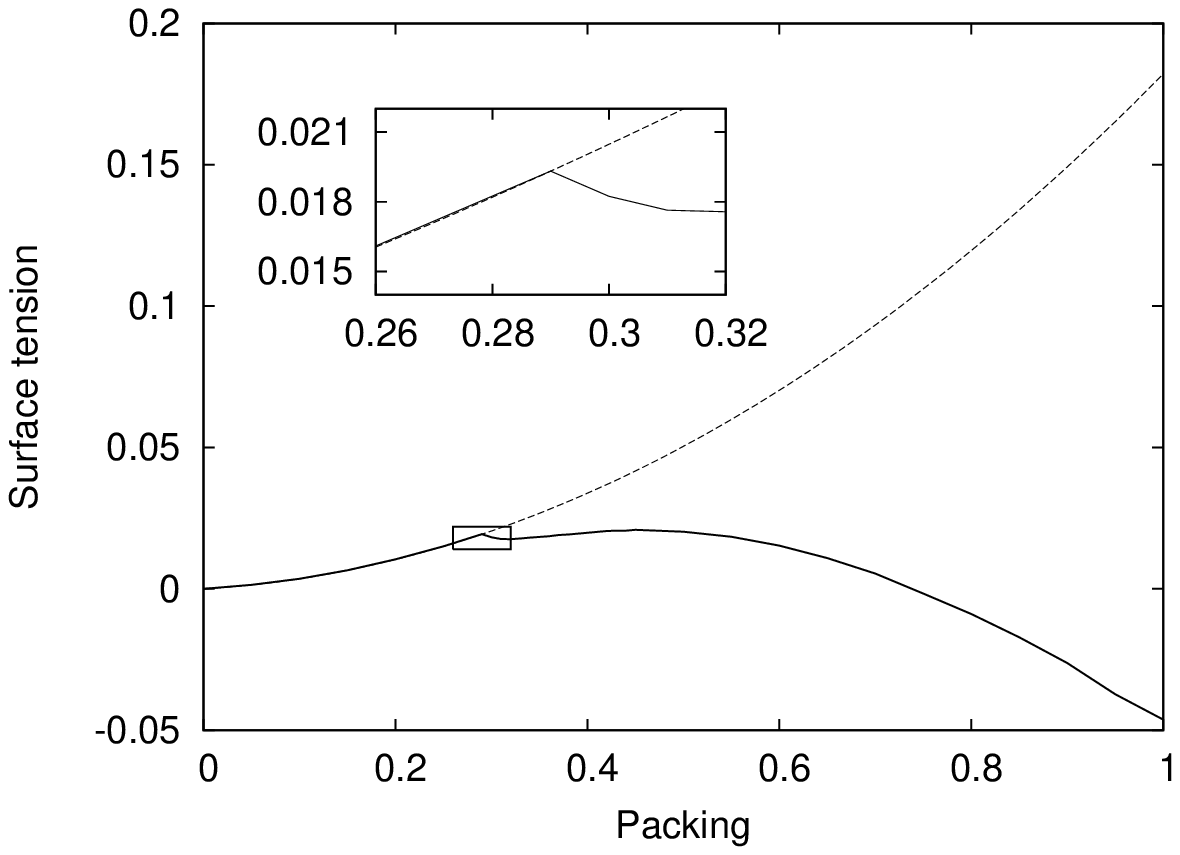}
\end{center}
\caption[Surface tension vs packing for system B).]{
\label{fig7}
Surface tension $\beta \gamma a^2$ vs packing (system B).
The dashed line describes values calculated for the isotropic phase.
The inset shows the neighborhood of the point $y_C=0.29437$
with the transition from the isotropic phase to the biaxial nematic phase.}
\end{figure}

The density and the order parameters profiles for system A,
for the biaxial nematic phase in the bulk,
are plotted in Fig.~\ref{fig5}.
The biaxiality is present also in the interfacial region.
The surface tension for systems A and B is shown
in Figs. \ref{fig6} and \ref{fig7}, respectively.
On increasing density, the surface tension increases, and
there is the maximum at the transition (in the bulk) from the isotropic
to the biaxial nematic phase.

\begin{figure}
\begin{center}
\includegraphics[width=0.5\textwidth]{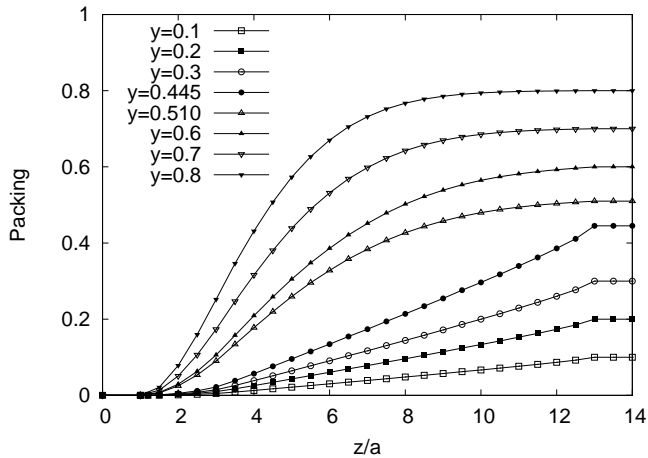}
\end{center}
\caption{
\label{fig8}
Comparison of the packing profiles $y(z)$ in the interfacial region
for different phase packing in the bulk (system A).}
\end{figure}

For high density, the surface tension decreases, but it can be
attributed to the fact that the low-density approximation
is no longer valid and the order parameters 
$\langle F_{\mu\nu}^{(j)} \rangle$ with $j>2$ are needed. 
Note that the closest packing of spheroplatelets in both systems 
is greater than 0.9.
The density profiles in the interfacial region for system A
are shown in Fig.~\ref{fig8}.

\begin{figure}
\begin{center}
\includegraphics[width=0.5\textwidth]{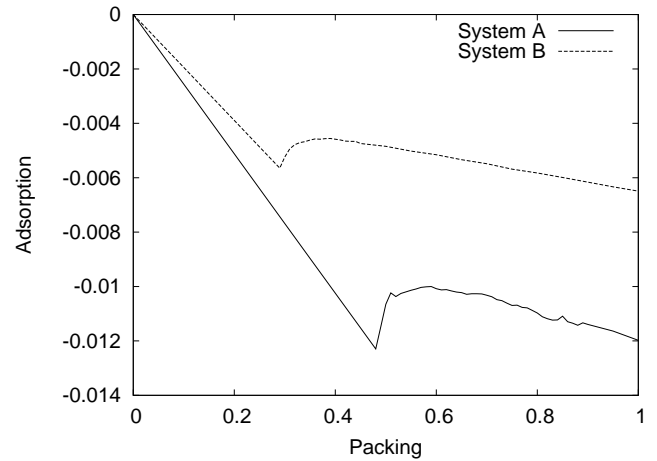}
\end{center}
\caption{
\label{fig9}
Adsorption $\Gamma a^2$ vs packing for systems A and B.}
\end{figure}

The density dependence of the adsorption is shown in Fig.~\ref{fig9}.
In the case of the isotropic phase, the adsorption decreases
according to a simple linear formula
$\Gamma = -\rho_0 ( a + b/4 + c/4 )$.
After the transition to the biaxial nematic phase the adsorption
first increases and then again decreases.
The adsorption is finite, and this suggests lack of the wall wetting
\cite{2007_Reich}.

\begin{table}
\caption{
\label{tab1}
Table reporting the values of the coefficients $L^{[I][J]}$
in $a^4$ units for system A. 
Errors estimated are less then 10\%.
The dagger symbol ($\dagger$) points to values that probably go to zero
(according to our tests).
}
\begin{center}
\begin{tabular}{cccccc}
\hline\hline
$L^{[I][J]}/a^4$ & $[J]=[1]$ & $[J]=[2]$ & $[J]=[3]$ & $[J]=[4]$ & $[J]=[5]$
\\ \hline
$[I]=[1]$  & 26209  & 7294  & -112  & 5.16 $\dagger$ & -4.97$\dagger$ \\
$[I]=[2]$  & -3415  & -1027  & 214  & -0.247$\dagger$  & 0.256$\dagger$ \\
$[I]=[3]$  & 1211  & 415  & -84.4  & 0.061 $\dagger$ & 0.214 $\dagger$\\ 
$[I]=[4]$  & 11.4$\dagger$  & 3.68 $\dagger$ & -0.622 $\dagger$ & -213  & 355 \\
$[I]=[5]$  & -3.16$\dagger$  & -0.921$\dagger$  & 0.154 $\dagger$ & 85.5  & -94.0 \\
\hline\hline
\end{tabular}
\end{center}
\end{table}

\begin{table}
\caption{
\label{tab2}
Table reporting the values of the coefficients $L^{[I][J]}$
in $a^4$ units for system B.   
Errors estimated are less then 10\%.
The dagger symbol ($\dagger$) points to values that probably go to zero
(according to our tests).
}
\begin{center}
\begin{tabular}{cccccc}
\hline\hline
$L^{[I][J]}/a^4$ & $[J]=[1]$ & $[J]=[2]$ & $[J]=[3]$ & $[J]=[4]$ & $[J]=[5]$
\\ \hline
$[I]=[1]$  & 110826  & 32921  & -4264  & 14.5$\dagger$  & -13.9 $\dagger$\\
$[I]=[2]$  & -15975  & -5075  & 960  & -0.317$\dagger$  & 0.002 $\dagger$\\
$[I]=[3]$  & 5355  & 2149  & -370  & -0.094  $\dagger$& -0.996 $\dagger$\\ 
$[I]=[4]$  & 36.0 $\dagger$ & 13.2 $\dagger$ & -1.4  $\dagger$& -1026  & 1911 \\
$[I]=[5]$  & -5.63 $\dagger$ & -2.55 $\dagger$ & 0.02$\dagger$  & 432  & -456 \\
\hline\hline
\end{tabular}
\end{center}
\end{table}

\section{Summary}
\label{sec:summary}

In this paper, we presented the statistical theory of hard molecules
near a hard wall in the low-density Onsager approximation.
A simple local approximation for the one-particle distribution
function was applied. The theory was used to study two systems
composed of the most biaxial hard spheroplatelets,
where the direct transition from the isotropic phase to the biaxial nematic
phase occurs in the bulk. The density and the order-parameter profiles 
near the wall were calculated.

The main result is the description of the phase near a wall
at the transition from the isotropic to the biaxial nematic phase.
Analytical results for the surface tension and the entropy 
contributions were presented.
The results should not depend on the low-density approximation
because they are the same for systems
with different molecule elongations.
The preferred orientation $(\vec{L},\vec{M},\vec{N})$ 
of the biaxial nematic phase is described by the condition
$\vec{L}=\vec{e}_z$, where the short molecule axes tend to be
perpendicular to the wall. 
The uniaxial symmetry along the axis perpendicular to the wall
must be broken spontaneously in order to set the vectors
$\vec{M}$ and $\vec{N}$.
The phase orientation imposed by the wall extends into the bulk
via the elastic forces.

For the case of the isotropic phase in the bulk,
the phase near the wall is uniaxial because some orientations 
are excluded by the presence of the wall.
The density profile of the phase in the interfacial region changes 
at the transition. If the phase is biaxial in the bulk
then more molecules can enter the interfacial region.
The complete wetting of the wall by a nematic film is not expected
because the transition from the isotropic to the biaxial nematic
phase is second order and the adsorption remains finite.

In order to confirm our predictions computer simulations 
of hard spheroplatelets near the wall are needed.
It would be interesting to check the density profiles 
in the interfacial region.
The validity of the local approximation for the one-particle
distribution function could be also tested.
We expect the short range density oscillations close to the wall.

Another interesting problem is the behavior of the system
composed of less biaxial molecules where, on increasing the density,
the following sequence of transitions is present:
the first-order transition from the isotropic phase to the uniaxial
nematic phase and, next, the second-order transition to the biaxial
nematic phase.

\begin{acknowledgments}

The authors are grateful to J. Spa{\l}ek for his support.
M. A. was supported by the Foundation for Polish Science (FNP) 
through Grant TEAM.

\end{acknowledgments}

\end{document}